\documentclass{PoS}

\title{Heavy-flavour production measurements in pp collisions at the LHC with ALICE}

\ShortTitle{Heavy-flavour production measurements in pp collisions at the LHC with ALICE}

\author{\speaker{Yvonne PACHMAYER} ~~for the ALICE Collaboration \\ %
        University of Heidelberg\\
        E-mail: \email{pachmay@physi.uni-heidelberg.de}}

\abstract{\begin{normalsize} The measurement of heavy-flavour production in pp collisions at the LHC allows to study
the production mechanisms and to test perturbative Quantum Chromodynamics (pQCD) in a new energy
domain. Further it provides important reference data for investigations of the medium effects in Pb-Pb
collisions, where charm and beauty production measurements are regarded as a unique probe for parton-medium interaction dynamics. We present preliminary results of open heavy-flavour production measurements with ALICE using hadronic D meson decays as well as semi-leptonic decays of D and B mesons and compare these with pQCD predictions. The dependence of $\rm J/\psi$ production on event multiplicity is also shown.\end{normalsize}}

\FullConference{XXIst International Europhysics Conference on High Energy Physics\\
		 21-27 July 2011\\
		 Grenoble, Rhones Alpes France}

\begin{document}

\section{Introduction}
\noindent A Large Ion Collider Experiment (ALICE) is the dedicated detector setup to study all aspects of heavy ion collisions at the LHC. For the investigation of the de-confined medium produced in central nucleus-nucleus collisions at high temperatures heavy-flavour production (open and hidden) is a particularly interesting probe because the heavy charm and bottom quarks experience the full
evolution of the collision. The measurement of heavy-flavour production in pp collisions provides the essential baseline to study production mechanisms and allows one to test pQCD calculations in a new energy domain. In the case of hadroproduction of quarkonium states non-perturbative aspects are also involved. \newline
\noindent The ALICE detector setup is well suited to detect and identify quarkonia as well as open charm and beauty hadrons due to a momentum resolution better than 2\% for $p_{\rm t}$ < 20 GeV/$\textit{c}$, a transverse impact parameter resolution better than 75(20) $\rm \mu$m for $p_{\rm t}$ > 1(20) GeV/$\textit{c}$ and various systems for particle identification, e.g. Time Projection Chamber (TPC), Time of Flight system (TOF), Transition Radiation Detector (TRD), Muon Spectrometer. The experiment and its heavy-quark detection performance are described in \mbox{detail in \cite{ALICEexperiment}.} \newline
Heavy-flavour production in pp collisions is measured in the following channels with ALICE:
\begin{itemize}
\item Open charm - fully reconstructed hadronic decays: \newline
\noindent $\rm D^{0} \rightarrow K \pi$, $\rm D^{0} \rightarrow K \pi\pi\pi$, $\rm D^{+} \rightarrow K \pi\pi$, $\rm D_{s} \rightarrow K K \pi$, $\rm D^{*} \rightarrow D^{0} \pi$, $\rm \Lambda_{c} \rightarrow \pi K p$ in $\rm |y| < 0.5$.
\item Open charm and beauty - reconstruction of leptons from semi-leptonic decay channels: \newline
\noindent D, B $\rm \rightarrow e + X$ in $\rm |y_{e}| < 0.8$ as well as D, B $\rm \rightarrow \mu + X$ in $\rm -4 < y_{\mu} < -2.5$; \newline 
\noindent displaced charmonia $\rm B \rightarrow J/\psi (\rightarrow e^{+}e^{-}) +X $ in $\rm |y| < 0.9$.
\item Quarkonia - $\rm c\overline{c}$ and $\rm b\overline{b}$ states in the dielectron ($\rm |y| < 0.9$) and dimuon ($\rm -4 < y < -2.5$) channel.
\end{itemize}

\section{Heavy-flavour production measurements in ALICE}
\begin{floatingfigure}[.4\textwidth]
\vspace{-0.4cm}
\includegraphics[width=.44\textwidth]{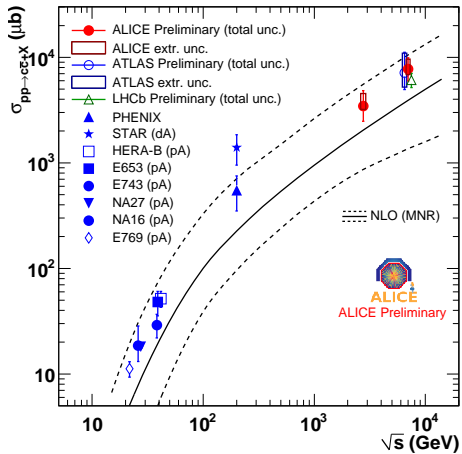}\caption{Total charm production cross section at various centre-of-mass energies.}\label{figure1}
\end{floatingfigure}

\noindent The reconstuction of open charm mesons is based on invariant mass analysis of its hadronic decay \cite{kschweda}.
The large combinatorial background was reduced by selecting displaced vertices\footnote{Separation of tracks originating from the secondary vertex of the decaying D meson from the collision vertex.} and additionally in the low transverse momentum region by identifying charged pions and kaons in the TPC and TOF. The correction for feed-down from B meson decays was done using FONLL predictions \cite{FONLLcalc}. So far, a pp data sample, with an integrated luminosity of $\rm L_{ int} = 1.6 ~nb^{-1}$ (1/3 of the total statistics in 2010), at a centre-of-mass energy of 7 TeV, has been analysed. This results in production cross section measurements covering the transverse momentum range of 2-12 GeV/$\textit{c}$ within the rapidity interval $\rm |y| < 0.5$. Predictions of pQCD calculations agree within uncertainties. \newline
\noindent The total $\rm c\overline{c}$ cross section was then determined from the production cross sections of $\rm D^{0}$, $D^{+}$, and $D^{*+}$, which were extrapolated to the full kinematic phase space and weighted according to the branching fractions measured at LEP. Details of the procedure along with the description of the uncertainty determination are published elsewhere \cite{kschweda}.  Figure \ref{figure1} shows the dependence of the total $\rm c\overline{c}$ production cross section on the centre-of-mass energy. To obtain the value at an energy of \mbox{$\sqrt{s}$ = 2.76 TeV} an analogous analysis was performed ($\rm L_{\rm int} = 1.1~nb^{-1}$). The corresponding results from ATLAS and LHCb were added to the plot \cite{ATLAS,LHCb}. The solid black line corresponds to next-to-leading-order predictions \cite{MNRcalc}, while the dashed lines depict its uncertainty. All data points lie at the upper edge of the predictions. \newline 

\begin{figure}[htbp]
  \centering
  \begin{minipage}{5 cm}
   \centering
    \includegraphics[width=1.4\textwidth]{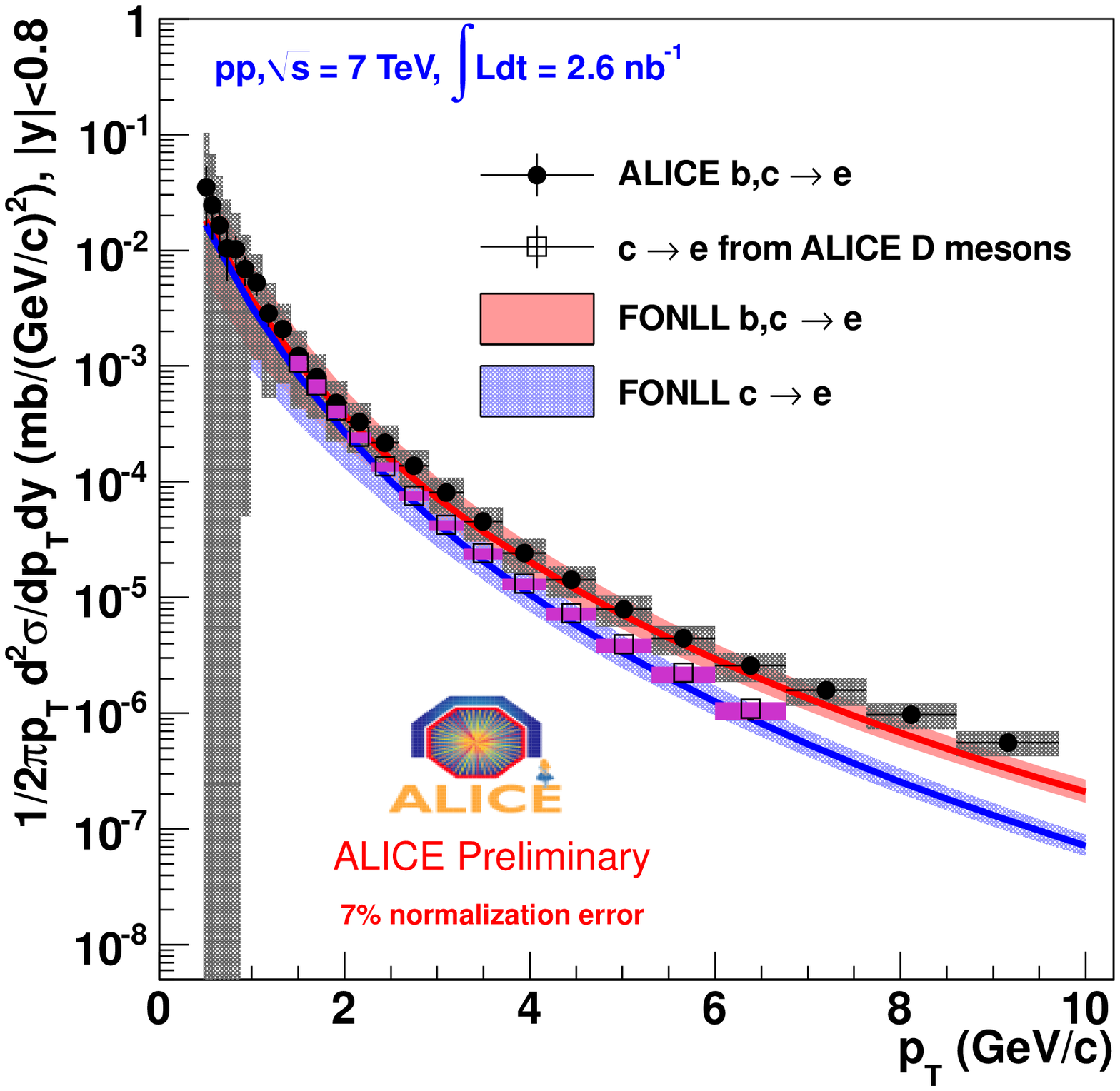} 
    \caption{Heavy-flavour decay electrons in $\rm |y| < 0.8$.}
    \label{figure2a}
  \end{minipage}
    \hspace{3.5cm}
  \begin{minipage}{5 cm}
   \centering
    \includegraphics[width=1.15\textwidth]{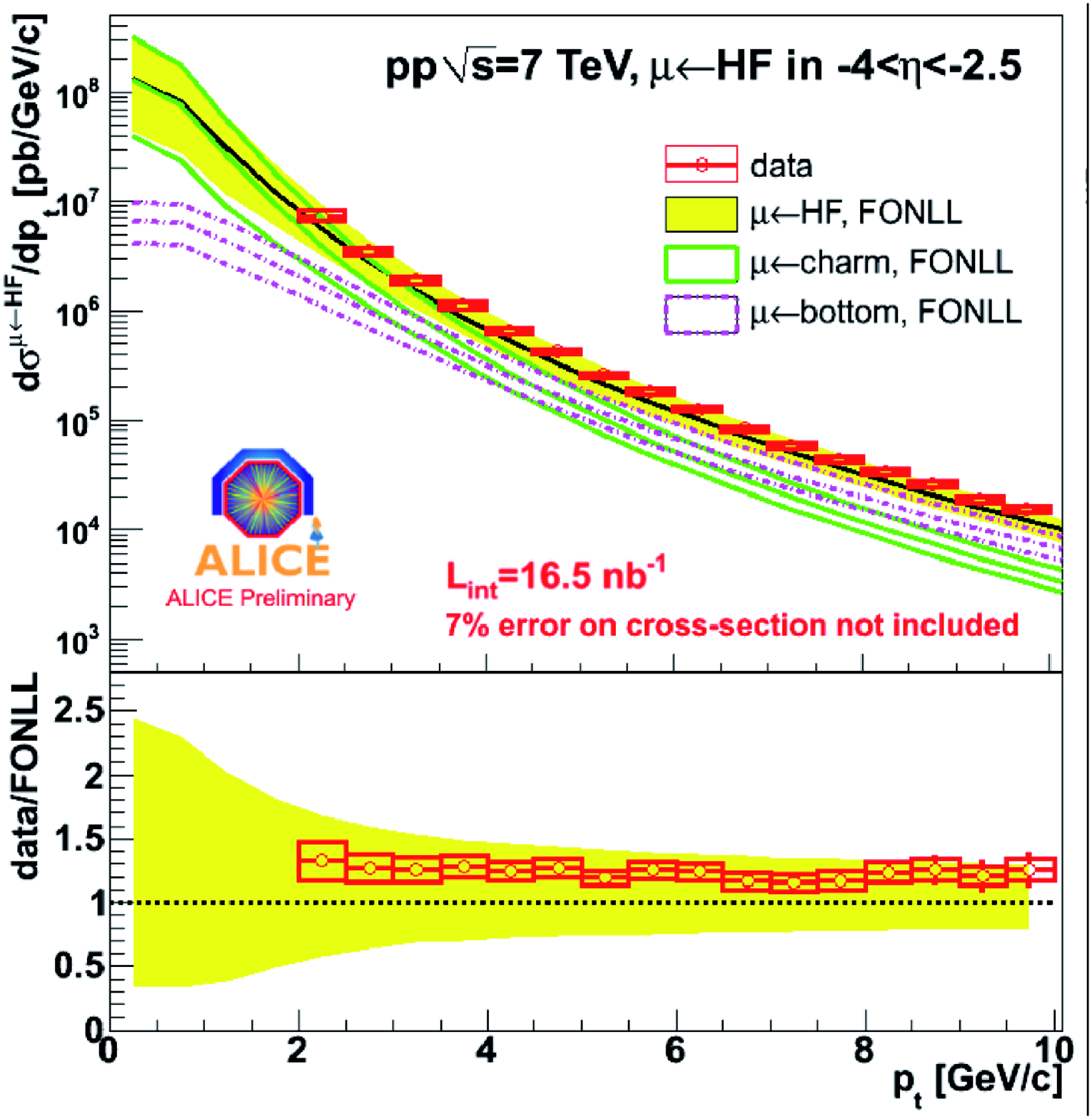}  
    \caption{Heavy-flavour decay muons in $\rm -4 < y < -2.5$.}
    \label{figure2b}
  \end{minipage}
\end{figure}

\noindent At mid-rapidity heavy-flavour production is also measured in the semi-electronic decay channel \cite{paperelectrons}. Electrons were identified using the signals in the TOF, the TPC and the TRD. The remaining hadron contamination was determined by performing a multiple Gaussian fit in momentum slices of the TPC dE/dx distribution, and the yield was subtracted from the electron spectrum. By requiring a hit in the innermost layer of the Inner Tracking System \mbox{(r = 3.9 cm)} the background from photon conversion was reduced. To extract the \mbox{$p_{\rm t}$ differential} production cross section of electrons from heavy-flavour hadron decays a cocktail of background electrons was subtracted from the inclusive electron spectrum. The cocktail was calculated \mbox{using} a Monte Carlo (MC) event generator and consists mainly of electrons from photon conversion and $\rm \pi^{0}$ Dalitz decays. Additionally electrons from decays of light mesons ($\rm \eta, \rho, \omega$ and $\rm \phi$), real and virtual direct photons as well as dielectron decays of $\rm J/\psi$ and Y are included. The $p_{\rm  t}$ spectra of those mesons were obtained based on the \mbox{$\rm \pi^{0}$ cross section}, from simulations and an $\rm m_{t}$-scaling ansatz \cite{paperelectrons}. The resulting $p_{\rm t}$ differential production cross section of electrons from heavy-flavour hadron decays within $\rm |y| < 0.8$ is shown in Fig. \ref{figure2a} ($\rm L_{\rm int} = 2.6~nb^{-1}$) in comparison with FONLL calculations \cite{FONLLcalc} and electrons from charm decays derived by applying the decay kinematics to the D meson cross sections. Using displaced vertex topology the contribution of beauty-decay electrons can be \mbox{disentangled \cite{paperelectrons}}. 

\noindent At forward rapidities ($\rm -4 < y < -2.5$) heavy-flavour production is measured in the semi-muonic decay channel by measuring single muons in the Muon Spectrometer and by removing the following background sources \cite{XZhang}: i) muons from the decay-in-flight of light hadrons, ii) muons from the decay of light hadrons produced in the interaction with the absorber (secondary muons) and iii) punch-through hadrons. The contributions of ii) and iii) were removed by matching reconstructed tracks with tracks in the muon trigger chambers \cite{ALICEexperiment}, applying a lower transverse momentum cut off ($p_{\rm t}^{\rm min} \rm = 2~GeV/\textit{c}$) and a DCA cut. The first contribution was subtracted from the inclusive spectrum by means of MC simulations. Figure \ref{figure2b} shows the resulting 
$p_{\rm t}$ differential production cross section of muons from heavy-flavour hadron decays ($\rm L_{\rm int} = 16.5~nb^{-1}$). The FONLL  \cite{FONLLcalc} prediction agrees with the data and indicates that beauty-decay muons dominate for $p_{\rm t} \rm > 6~GeV/\textit{c}$.\newline \newline
       \vspace{-0.6cm}
\begin{floatingfigure}[.45\textwidth]
\vspace{-0.4cm}
\includegraphics[width=.5\textwidth]{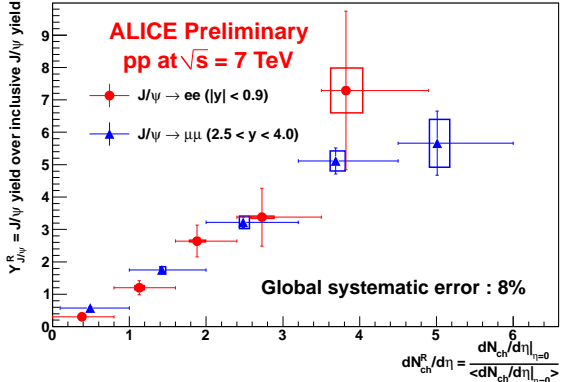}\caption{Relative $\rm J/\psi$ yield vs. relative charged particle multiplicity at forward and mid-rapidity.}\label{figure3}
\end{floatingfigure}

\noindent The inclusive $\rm J/\psi$ production measurement was based on the invariant mass analysis of two 
at mid-rapidity identified electrons \mbox{(using} signals of the TPC) and at forward rapidity identified muons $~$ \mbox{(Muon Spectrometer).} With both channels it is possible to study $J/\psi$ production over a large rapidity range $\rm |y| < 0.8$ and $\rm -4 < y < -2.5$  in both cases with an acceptance down to $p_{\rm t} = 0$. Details on the analysis as well as $p_{\rm t}$- and y-differential production cross sections can be found in \cite{JPsi}.\newline
\noindent Figure \ref{figure3} shows the relative $\rm J/\psi$ yield vs relative charged particle multiplicity in the di-electron ($\rm L_{\rm int} $ $\rm = 5.6~nb^{-1}$) and di-muon ($\rm L_{\rm int} = 15.6~nb^{-1}$) channel for pp collisions at $\rm \sqrt{s} = 7 TeV$. The relative yield is defined as the per-event yield in each multiplicity bin, normalized by the inclusive yield per pp inelastic event. The relative charged particle multiplicity describes the charged particle multiplicity normalized to the respective minimum bias value. An almost linear dependence is visible for both cases. This is the first measurement of $\rm J/\psi$ vs multiplicity and the theoretical interpretation is still lacking.

\vspace{-0.3cm}
\section{Conclusion and outlook}
\vspace{-0.3cm}
\noindent The measurements of heavy-flavour production cross sections in pp collisions at $\sqrt{s}$ = 7 TeV with ALICE have been presented. FONLL calculations agree within uncertainties with our open heavy-flavour results at forward and mid-rapdidity, but the data are rather on the high side of the calculations. Using the full statistics collected in 2010 the transverse momentum range of the full reconstructed charm mesons will extend down to 1 GeV/$\textit{c}$ and up to at least 20 GeV/$\textit{c}$. This high-statistics data sample will allow to study $\rm D_{s}$ and $\rm \Lambda_{c}$ production as well. The $\rm J/\psi$ production measurement is performed down to zero $\rm p_{\rm t}$ at both forward and mid-rapidity. With increasing event multiplicity the number of produced $\rm J/\psi$ mesons increases. Polarization studies as well as an investigation to separate prompt and secondary $\rm J/\psi$ are ongoing.

\end{document}